\documentclass{appolb}
\usepackage{epsfig}

\newcommand{\boldtau}{\mbox{\boldmath $\tau$}}
\newcommand{\boldpi}{\mbox{\boldmath $\pi$}}

\begin{document}
\pagestyle{plain}
\newcount\eLiNe\eLiNe=\inputlineno\advance\eLiNe by -1
\title{Charge symmetry breaking in pion production}
\author{J. A. Niskanen\thanks{Email Jouni.Niskanen@helsinki.fi}
\address{Department of Physical Sciences, P. O. Box 64 \\
FIN-00014 University of Helsinki, Finland}}

\maketitle

\begin{abstract}
Chiral effective field theory predicts a charge symmetry violating
(CSB) amplitude for pion-nucleon scattering. This mechanism provides
a very large contribution also to the CSB forward-backward asymmetry
in the angular distribution of the reaction $pn\to d\pi^0$.
This contribution was so large that it had a
potential to cause a large effect also in CSB elastic
$NN$ scattering and to disturb its present understanding.
However, it can be seen that, contrary to pion
production, in this case the $ud$-quark mass difference
and electromagnetic contribution to the $np$-mass
difference tend to cancel causing the total effect in
the effective range parameters $\Delta a = a_{pp} - a_{nn}$
and $\Delta r_0 = r_{0,pp} - r_{0,nn}$ to
be relatively small. In the lowest order and within the static
approximation for the nucleons CSB pion-nucleon
rescattering does not influence $np$ scattering.
\end{abstract}

\section{Introduction}

Charge symmetry is a special case of the general flavour
symmetry of QCD, which at its simplest distinguishes
the proton and the neutron (or $u$ and $d$ quarks).
It is, of course,
trivially broken by the electromagnetic interaction, notably the
Coulomb force in comparisons of the $pp$ and $nn$ systems and by
the magnetic interaction in the $np$ system. Other
well known sources are the $np$ mass difference and $\eta\pi$- as well
as $\rho\omega$-meson mixing. These in turn may be related to the
up- and down-quark mass difference - the microscopic flavour symmetry
breaking in QCD. One might consider remarkable the fact that, although
the relative quark mass difference is large ($\ge 10\%$), the symmetry
breaking at the observable hadron level is two orders of magnitude
smaller.

In the mirror system $pp$ vs. $nn$
CSB has been studied for many decades \cite{physrep},
while its appearance in the $np$ system was first seen only a decade
ago \cite{classiv} as the difference $\Delta A = A_n - A_p$ of elastic
analyzing powers.  Different CSB observables
have been seen in calculations
to be sensitive to different combinations of sources. For example, in
$np$ scattering above 300 MeV the $np$ mass difference in OPE dominates,
while at and below
$\approx$ 200 MeV $\rho\omega$ meson mixing and the magnetic
interaction become about equally important\cite{wtm}. In contrast,
the $pp$ vs. $nn$ difference is dominated by $\rho\omega$ meson
mixing \cite{physrep,tera,cb}. The CSB effects in the $np$ system
change the total isospin of the two nucleons
(class IV in the terminology of Ref. \cite{henley}), whereas in
$pp$ and $nn$ the isospin must be conserved (class III).

\section{CSB pion production}

It is easily seen that in the reaction $np \rightarrow d\pi^0$
the isospin change $T=0\rightarrow 1 $ implies a CSB asymmetry of
the unpolarized cross section about 90$^\circ$ and vice versa.
Namely, due to the generalized Pauli principle and conservation
of the angular momentum and parity, with isospin one initial
states for odd $l_\pi$ only singlet-even initial
states are possible and for even $l_\pi$ only triplet-odd states.
The presence of some isospin zero component in the
initial state introduces  opposite
spin-parity assignments and the initial spin states will
then have both parities involved.
This minor asymmetry is being measured in an on-going
experiment at TRIUMF \cite{e704}.

Since class IV forces mentioned above change the isospin,
as an initial state interaction they can
quite naturally give rise to $d\pi^0$ final states even from initial
$T=0$ $np$ states. Further, although in a purely nucleonic
basis $\eta\pi$ mixing force conserves the isospin (class III),
it can cause a $T=0$
$\Delta N$ admixture even in initial isospin
zero states \cite{few,nst,meson2000} and thus contribute
to pion production from these states. In addition $\eta\pi$ mixing
can contribute very explicitly
in the final actual production vertex in the form of production first
of an isoscalar $\eta$ meson which then transforms
into a pion.

Of traditional
CSB mechanisms in pion production $\eta\pi$
mixing is important and was seen to dominate at threshold \cite{few},
while the $np$ mass difference becomes more
important at higher energies, where the two $\eta\pi$ mixing
mechanisms described above tend to cancel \cite{nst}.

\begin{figure}[tb]
\begin{center}
\epsfig{file=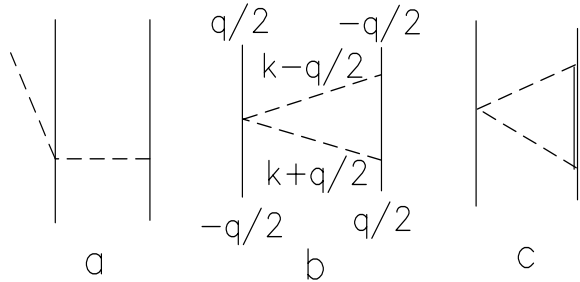,width=4cm,height=8cm,angle=90}
\end{center}
\caption{Isospin breaking
pion-nucleon rescattering mechanisms in $np \rightarrow d\pi^0$
(a) and in $NN$ elastic scattering (b and c).}
\end{figure}

Exploiting the fact that  chiral symmetry gives predictions for effects
arising from the small but explicit breaking of this symmetry
generated by the quark masses, Ref. \cite{knm} employed
a CSB effective Lagrangian based on
chiral symmetry and including the $u$ and $d$ quark mass differences
\begin{equation}
 {\cal L}_{\rm qm}^{(1)}=
 \frac{\delta m_{N}}{2}
                       \left(N^\dagger \tau_{0}N
                        -\frac{2}{DF_{\pi}^{2}}
   N^\dagger \pi_{0} \boldpi \cdot \boldtau N \right)
\end{equation}
to describe pion-nucleon $s$-wave rescattering in $np\rightarrow d\pi^0$.
This rescattering dominates isospin conserving production at threshold.
Here $\boldtau$ represents the Pauli matrices in isospin space,
$F_\pi=186$ MeV is the pion decay constant, and
$D=1+\boldpi^2/F_\pi^2$, though for simplicity $D=1$ was used.
It is important to note that the pion-nucleon interaction in
the second term is linked by chiral symmetry to the first term, which
in turn is directly related to the neutron-proton mass difference
giving $m_n-m_p=\delta m_{N}+\bar\delta m_{N}$. Here $\bar\delta m_N$
is an electromagnetic "hard photon" exchange contribution to
the neutron-proton mass difference also related
to the isospin violating pion-nucleon interaction in
the effective Lagrangian
\begin{equation}
 {\cal L}_{\rm hp}^{(-1)} =
 \frac{\bar{\delta} m_{N}}{2}
                       \left(N^\dagger \tau_{0}N
                        +\frac{2}{DF_{\pi}^{2}}
      N^\dagger (\pi_{0}
      \boldpi\cdot \boldtau -\boldpi^2 \tau_0) N\right).
\end{equation}
This CSB rescattering mechanism is presented in Fig. 1a.

Above,  the quark mass difference is by no means small as
compared with their sum but rather
 $m_d-m_u \equiv \varepsilon (m_d+m_u)$
with $\varepsilon \sim 1/3$, although the individual parameters
$\delta m_N$ and $\bar\delta m_N$ are not completely
uniquely determined and remain model dependent.
Their values have reasonable ranges
2--3 MeV and (--0.5)--(--1.5) MeV, respectively.

At the TRIUMF experiment energy 279.5 MeV
the integrated forward-backward asymmetry
\begin{equation}
A_{fb}=\frac{\int_0^{\pi/2}
           d\Omega \; [\sigma(\theta)-\sigma(\pi-\theta)]}
{\int_0^\pi d\Omega \; \sigma(\theta)}.
              \label{asym}
\end{equation}
was obtained in Ref. \cite{knm} in terms of the above mass
differences as
\begin{equation}
A_{fb} \simeq \left(-28
  +{24\over {\rm MeV}}
                     (\delta m_N -\frac{\bar\delta m_N}{2})
               \right) \times 10^{-4} ,\label{afb}
\end{equation}
where the conventional meson contributions in the first term have been
taken from Ref. \cite{few}. Due to the dominance of the $\eta\pi$
mixing, the uncertainties in its strength directly scale that
term. Apart from minor pionic effects due to the $np$ mass
difference, it is proportional to $G_\eta \langle\pi |H|\eta\rangle$.
The values $G_\eta^2/4\pi = 3.68$ and $\langle\pi |H|\eta\rangle
-5900$ MeV$^2$ have been used for this result. (In addition, the
analogous effect from $\eta'$ with the same coupling and the mixing
matrix element --5500 MeV$^2$ was icluded \cite{few}.)

Since in Eq. (\ref{afb}) $\delta m_N$ and
$\bar\delta m_N$ add constructively, the latter term arising
from the new effective Lagrangian easily dominates the total $A_{fb}$
making it change sign. With the given first term, the
total $A_{fb}$ would vary in the range 0.3--0.6\%. It is
unlikely that the first term is an overestimate, so the
range might rather be even more positive. With the anticipated
experimental resolution of 0.12\% \cite{e704} this is a significant
result. If the uncertainties in the $\eta\pi$ mechanisms (notably
the $\eta  NN$ coupling constant) can be solved, CSB pion
production could set constraints also on $\delta m_N$ and
$\bar\delta m_N$.

\section{Effect in elastic $NN$ scattering}

Since the new effective CSB interaction gave such a large
contribution in pion production, one is prone to worry how
much this $\pi N$ rescattering might contribute to elastic
$NN$ scattering possibly breaking the consensus there. In Ref.
\cite{elas} the CSB $NN$ two-pion exchange interaction
depicted in Fig. 1b was
shown in lowest order (in the static approximation for baryons)
to be of the form of class III
\begin{equation}
V_N(q) =
 \frac{\delta m_{N}+2\bar\delta m_{N}}{F_\pi^2} \frac{f^2}{\mu^2}
\int \frac{d^3k}{(2\pi)^3} \frac{(k^2-q^2/4)(\tau_{10}+\tau_{20})}
{[\mu^2+({\bf k} +{\bf q}/2)^2] [\mu^2+({\bf k} -{\bf q}/2)^2]},
\end{equation}
where $f^2/4\pi = 0.076$ is the pion-nucleon coupling constant
and $\mu$ the pion mass. A similar expression resulted for
exchanges involving an intermediate $\Delta(1232)$ isobar
(Fig. 1c).

Contrary to pion production, in this case,
although the integrals and parameters are
rather large, the two mass differences cancel to a large extent
leaving only rather small contributions to the
difference between $nn$ and $pp$ scattering. To see
this, one need only compare the range of the factor
$\delta m_{N}+2\bar\delta m_{N}$ [+0.8--(-0.2) MeV] with
the range of $\delta m_N -\bar\delta m_N/2$ [2.2--3.4 MeV].
Correspondingly, the contributions to the scattering
length and effective range differences  $\Delta a = a_{pp} - a_{nn}$
and $\Delta r_0 = r_{0,pp} - r_{0,nn}$ vary between
0.17 fm and --0.05 fm, and 0.003 and --0.001 fm, respectively.
Also the contribution from this source to the $^3$H--$^3$He
binding energy difference is expected to be small.


\begin{thebibliography}{99}
\bibitem{physrep} G. A. Miller, B. M. K. Nefkens and I. Slaus,
Phys. Rep. {\bf 194}, 1 (1990).
\bibitem{classiv}
R. Abegg {\it et al.}, Phys. Rev. Lett. {\bf 56}, 2571 (1986),
Phys. Rev. D {\bf 39}, 2464 (1989);
 L. D. Knutson {\it et al.}, Phys. Rev. Lett. {\bf 66}, 1410 (1991);
 S. E. Vigdor {\it et al.}, Phys. Rev. C {\bf 46}, 410 (1992);
 J. Zhao {\it et al.}, Phys. Rev. C {\bf 57},  2126 (1998).
\bibitem{wtm} A. G. Williams, A. W. Thomas and G. A. Miller,
Phys. Rev. C {\bf 36}, 1956 (1987); M. J. Iqbal and J. A. Niskanen,
Phys. Rev. C {\bf 38}, 2259 (1988).
\bibitem{cb} S. A. Coon and R. C. Barrett, Phys. Rev. C {\bf 36},
2189 (1987).
\bibitem{tera} G. F. de T\'eramond, Proc. Symposium/Workshop on Spin
and Symmetries, TRIUMF, Vancouver, June 30--July 2, 1989, TRI-89-5,
eds. W. D. Ramsay and W. T. H. van Oers, p. 235.
\bibitem{henley} E. M. Henley and G. A. Miller, in {\it Mesons and
Nuclei}, Vol. I, eds. M. Rho and D. H. Wilkinson, North Holland
(Amsterdam 1979).
\bibitem{e704} TRIUMF experiment E704, spokespersons A. Opper and
E. Korkmaz.
\bibitem{few} J. A. Niskanen, Few-Body Systems {\bf 26}, 241 (1999).
\bibitem{nst} J. A. Niskanen, M. Sebestyen, and A. W. Thomas,
Phys. Rev. C {\bf 38}, 838 (1988).
\bibitem{meson2000} J. A. Niskanen,  Proc. of Meson2000 Workshop,
Cracow, May 2000, Acta Physica Polonica {\bf B 31}, 2683 (2000).
\bibitem{knm} U. van Kolck, J. A. Niskanen and G. A. Miller,
Phys. Lett. B 493, 65 (2000); nucl-th/0006042.
\bibitem{elas} J. A. Niskanen,  Phys. Rev. C {\bf 65}, 037001 (2002);
nucl-th/0108015.

\end{thebibliography}
\end{document}